\documentclass[%
reprint,
amsmath,amssymb,
aps,
prd,
longbibliography,
lengthcheck,%
]{revtex4-1}

\usepackage{hyperref}

\begin{document}

\preprint{CERN-PH-TH/2009-242}

\title{New Renormalization Group Equations and the Naturalness Problem}

\author{Grigorii Pivovarov}
\affiliation{Institute for Nuclear Research of Russian Academy of Sciences, 117312 Moscow, Russia}

\date{\today}

\begin{abstract}

Looking for an observable manifestation of the so-called unnaturalness of scalar fields we introduce
a seemingly new set of differential equations for connected Green functions. These equations describe the 
momentum dependence of the Green functions and are close relatives to the previously known renormalization 
group equations. Applying the new equations to the theory of scalar field with $\phi^4$ interaction we 
identify a relation between the four-point Green function and the propagator which expresses the 
unnaturalness of the scalar field. Possible manifestations of the unnaturalness at low momenta 
are briefly discussed.

\end{abstract}


\maketitle

\section{Introduction}

It was noted in \cite{Wilson:1970ag, *'tHooft:1979bh} that scalar fields have ultraviolet properties that set them apart from the fields 
with nonzero spin. Namely, the scalar mass term does not break any symmetry (in contrast to the cases of spinor and 
gauge vector fields). Therefore, quantum corrections to scalar mass have not to vanish in the limit of zero mass, 
and may be proportional to cutoff. Direct computations demonstrate that this is indeed the case for particular models 
(for example, this is the case for the standard model). In \cite{Susskind:1978ms} it was argued that such a situation 
is physically inappropriate. The difficulty was dubbed ``naturalness problem''. It may be formulated in the following 
way: physical measurements at a moderate energy restrict the measurements of the same parameters at a higher energy, 
and the restrictions become stronger with growing energy. In other words, to see the world the way we see it, the 
effective theory defined at a high energy should be fine tuned, which seems to be unnatural.

We are interested in a precise formulation of this problem: which observables should be tuned and to what precision?
The original reasoning argues that bare parameters of the model should be tuned to a precision which grows as a power 
of the cutoff \cite{Susskind:1978ms}. Loosely speaking, bare parameters at a given cutoff are amplitudes of certain 
reactions taken at the energy scale of the cutoff. For example, bare mass at a cutoff $\Lambda$ may be related to a 
scalar propagator taken at the Euclidean momentum squared $\Lambda^2$, and the bare scalar self-coupling may be 
related to a two-by-two scattering amplitude of the scalar particles at the same energy scale. So, we expect that 
the unnaturalness of the scalar fields can be expressed as unnatural relations between the renormalized Green 
functions of the theory.

Until now, the studies in this field were mostly concerned with the matters of principle. We want to attempt here a 
more phenomenological attitude: even if in a fundamental theory some mechanism cures the unnaturalness of scalar fields,  
remnants of the unnaturalness may be seen at low energies. For example, unnatural relations between the amplitudes 
involving scalar fields may still hold approximately at low energies. To our knowledge, no such relations have been 
suggested. Moreover, the most popular regularization, viz. dimensional regularization, does not employ any cutoff 
and does not have quadratic divergences, and it was claimed on these grounds that the naturalness problem is 
inexistent \cite{Bardeen:1995kv}. On the other hand, it was pointed out that quadratic divergences show up within 
dimensional regularization as poles in the dimension when it approaches rational values $4-2/m$, where $m$ 
is the number of loops \cite{Veltman:1980mj}.

We derive a relation between the running scalar mass and the scalar self-coupling renormalized with momentum 
subtractions. This relation holds approximately for a range of high normalization points. We derive this 
relation within $\phi^4$ theory. (The range of values of the normalization points for which our relation 
holds is discussed below.)
The relation reads as follows:
\begin{equation}
\label{result}
M^2(Q^2) = \gamma_\phi Q^2 + \dots.
\end{equation}  
Here $Q^2$ is the normalization point, $\gamma_\phi = g^2/(12(16 \pi^2)^2)$ is the anomalous dimension 
of the scalar field within $\phi^4$ theory, and dots denote the higher order corrections in the expansion 
of the right-hand-side in powers of the coupling. We should stress that the appearance of the anomalous 
dimension in this equation is not understood by the present author at the moment. For example, as of now, 
we do not claim that this relation holds in arbitrary renormalizable model involving scalar field. 
Such transient unexpected relations between amplitudes involving scalar fields, if they exist within 
the standard model, may be of importance for describing the scattering of the longitudinal vector bosons, 
where some kind of resummation is welcome \cite{Veltman:1993pr}. 

The above relation between the running mass and coupling expresses the unnaturalness of scalar fields. 
(The unnaturalness consists in appearance at high energy of a relation between the parameters that are 
independent at low energy.) It is an outcome of an evolution equation for the renormalized scalar propagator 
governing the dependence of the propagator on the momentum squared. This equation belongs to a set of 
equations (we derive the whole set) expressing independence of the Green functions of the normalization 
points. It is an equation of second order in derivatives over momentum squared. The appearance of the 
second derivatives over momentum squared is in accord with the presence of the quadratic divergences 
in the theory. The solution to the equation effectively resums perturbative corrections.

The latter can also be stated about the MS-scheme renormalization group equations 
\cite{Collins:1973yy}---the version of the renormalization group equations which is most popular 
lately. A comparison is in order. Let us make it on the example of $\phi^4$. For this model, 
effectively, the MS-scheme renormalization group resummation leads to the following transformation 
of the free scalar propagator (the mass is neglected):
\begin{equation}
\label{ms}
\frac{1}{Q^2}\rightarrow\frac{1}{Q^{2(1-\gamma_\phi)}\mu^{2\gamma_\phi}}.
\end{equation}
Here $\gamma_\phi = g^2/(12(16 \pi^2)^2)$ is the scalar field anomalous dimension \cite{Collins:1974ce}, 
$g$ is the coupling renormalized within MS-scheme, $\mu$ is the MS-scheme mass unit, and we neglected 
the corrections related to the running of the coupling. As we see, summation of the corrections of 
the form $Q^2(g^2 \log Q^2)^n$ results in a softening of the propagator. Because of this, subleading 
corrections to the scalar self-energy proportional to $g^m Q^2$ may be numerically important. Indeed, 
at large momentum, $Q^2 \gg Q^{2(1-\gamma_\phi)}\mu^{2\gamma_\phi}$. It seems that the MS-scheme 
renormalization group ignores this kind of corrections. At the same time, they are certainly present 
withing the momentum subtraction scheme. The simplest way to see it is to take a second derivative 
in momentum squared of the sunset diagram. This derivative is a convergent Feynman integral. It can 
be analytically computed in the zero mass limit  \cite{Groote:2005ay}. At infinite momentum, the 
contribution of the sunset diagram to the scalar self-energy can be computed as a double integral 
of the above second derivative taken at zero mass. The outcome of this computation is
\begin{equation}
\label{ss}
\Sigma(Q^2) \approx \gamma_\phi \big(Q^2\log\frac{Q^2}{Q_0^2} - (Q^2-Q_0^2)\big),
\end{equation}
where $\Sigma$ is the self-energy, $Q_0$ is the normalization point where the correction vanishes. 
The term involving $\log$ is treated within the MS-scheme renormalization group. The term without 
the $\log$ is not treated within the MS-scheme renormalization group, but both terms are treated 
within the new equations we suggest.

The interplay between the logarithms and powers of the momentum yields a reach evolution pattern. 
To describe it we employ a number of objects. The first one is the inverse propagator $R(Q^2)$. 
For the free theory, $R(Q^2) = Q^2 + M^2$  (throughout the paper we use the Euclidean formulation). 
Next is the running mass $m^2(Q^2)$ in units of the normalization point $Q^2$. By definition, 
$m^2(Q^2)\equiv R(Q^2)/\big(Q^2 R'(Q^2)\big) - 1$. For free theory, $m^2(Q^2) = M^2/Q^2$. One can 
quantify the deviation from the free theory checking the decay of $m^2(Q^2)$ at large momentum. 
Lastly, we use as an indicator of the evolution in momentum squared the second power of the wave-function 
normalization, $n(Q^2)\equiv (R'(Q^2))^4$. It is an important quantity, because its inverse appears 
in an effective expansion parameter of the perturbation theory in $g^2$. Thus, perturbation theory 
is unreliable when $n(Q^2)$ becomes small.

We will see that the second order evolution equation for the propagator can be recast as a pair 
of the first order coupled evolution equations for the above $m^2(Q^2)$ and $n(Q^2)$. It will 
also turn out that $n(Q^2)$ is slowly decreasing when $Q^2$ is growing, 
$n(Q^2) \approx 1 - 4\gamma_\phi\log(Q^2/M^2)$. If we ignore this logarithmic decrease, 
we will see that $m^2(Q^2)$ goes down until it takes the value $\gamma_\phi$, which happens at 
$Q^2 \approx M^2/\gamma_\phi$ (notice that $n(Q^2)$ is still close to unit for this value of 
$Q^2$ if $\gamma_\phi$ is small enough). At this value the decrease of $m^2(Q^2)$ stops and 
it stays approximately constant growing very slowly until the $\log$-term in the above expression 
for $n(Q^2)$ becomes comparable with unit. This sets up the range of applicability of the 
relation (\ref{result}): $M^2/\gamma_\phi < Q^2 \ll M^2 \exp(1/(4\gamma_\phi))$. The upper bound 
also sets up the range of applicability of the perturbation theory. If we go beyond the range 
of applicability of the perturbation theory in momentum squared, $m^2$ grows rather fast due to 
decrease of $n$, and the inverse propagator reaches the asymptotic value 
$R_a\equiv 2M^2 \exp(3/(2\gamma_\phi\times 0.3609))$ at $Q^2 = Q_c^2 \propto R_a/\gamma_\phi$, 
and the evolution stops (the numerical value in the exponent of $R_a$ is approximate). 
If we would trust this result, it would mean that beyond the critical value of the momentum, 
$Q_c$, the scalar propagator becomes local, $D(x) = \delta(x)/R_a$ in the coordinate space. 
We stress here that this last property is not a property of the scalar propagator, but only
a property of a solution to the leading order renormalization group equation for the scalar propagator
considered in the domain where the higer order corrections to the renormalization group equation 
may be important.

Our derivation of the new renormalization group equations is quite general. It is valid for 
any renormalizable field theory.  It rests on a definition of a convenient renormalization scheme. 
Details are given below (Sections 3 and 6). For now, it suffices to say that we fix the theory 
by fixing the values of the propagators and a sufficient number of their derivatives at a value 
of the momentum squared, and, on top of that, we fix values of a sufficient set of connected 
Green functions with external propagators amputated, and momenta set to some values. Clearly, 
this is almost the momentum subtraction (mom) scheme. It differs the mom scheme by the 
one-particle-reducible contributions to the quantities we choose for our normalization conditions. 

Apart of usefulness in deriving the new renormalization group equations, there is another reason 
for choosing this particular scheme. Namely, the quantities chosen for normalization are directly 
related to amplitudes (we normalize by fixing normalizations of amplitudes at points in the space 
of their variables). In fact, the new renormalization group equations describe momentum dependencies 
of the propagators and of the most desirable amplitudes of the theory with small numbers of involved particles. 

Generally, the approach we develop can be viewed as a partucular realization of the approach to renormalization advocated in
\cite{Lepage:1989hf}: In this approach choosing a renormalization scheme is identified with choosing a set of observables used to parameterize the theory. Furthermore, renormalizing is identified with expressing any other observable of the theory in terms of the observables involved in choosing the renormalization scheme. A theory is called renormalizable if the set of observables defining the renormalization scheme is finite, and after the renormalization the observables stay finite at infinite cutoff. We suggest a particular renormalization scheme within this approach. The set of observables defining our renormalization scheme is contained in the normalized action, which is a local functional of the fields of the model we introduce in Section 3.

The new renormalization group equations are easy to derive for this scheme because it is possible 
to use the Dominicis-Englert theorem \cite{deDominicis:1967px}. This theorem states that there 
is a duality between bare action and the generating functional of the renormalized connected Green 
functions: there are Feynman rules for expressing the latter in terms of the former, but the action 
can also be expressed in terms of the renormalized connected Green functions with similar Feynman rules. 
We take the expression of the action in terms of Green functions, subtract from it the local part and 
obtain zero because the bare action is local. The subtraction procedure involves the normalization points. 
In this way we obtain equations for Green functions depending on the normalization points. These 
equations do not involve the bare action. For that we call them ``inaction equations''. Next step 
demonstrates that inaction equations can be solved perturbatively for Green functions. In this way we 
set up a new perturbation theory which does not involve any divergent quantities. In particular, 
the new scheme does not involve the divergent bare action. Instead, the scheme involves a 
finite ``normalized action'' which is a specific combination of renormalized connected Green 
functions. All this constitutes a quite general scheme of constructing perturbation theory 
that avoids any encounter with divergent quantities. Such schemes have been discussed in the 
literature (see \cite{'tHooft:2004mn} and references therein). But our version seems to be more 
analytical, because we are able to express it in equations without referring to combinatorial 
manipulations with infinite sets of Feynman diagrams. 

The last step in derivation of the new renormalization group equations is to take the derivatives 
in normalization points of the inaction equations. It yields differential equations describing 
dependence of the normalized action on the normalization point.

Finally, we apply the above scheme to $\phi^4$ and obtain a differential equation for the 
inverse propagator, $R\equiv 1/D$. This equation has the following form:
\begin{equation}
\label{prop}
R'' = -\frac{8\gamma_\phi}{(R')^3Q^2}\int_0^\infty  J_3(x)[mK_1(mx)]^3 x \;dx + \dots,
\end{equation}
where $\gamma_\phi\equiv g^2/(12 (16\pi^2)^2)$, $m^2\equiv R/(R'Q^2) - 1$, the primes denote 
the derivatives in $Q^2$, the dots denote higher order corrections, $J_3$ is the Bessel 
function of the third order, and $K_1$ is the modified Bessel function of the first order. 
From this it is quite straightforward to deduce the above relation (\ref{result}). 

The paper is organized as follows. In the next Section we fix our notations and present a 
short derivation of the Dominicis-Englert theorem. In Section 3 we derive the inaction 
equations and introduce the normalized action. In Section 4 we work out perturbation 
theory in couplings, which are parameters of the normalized action. Also in this Section 
the first nontrivial order of the perturbation theory is described. In Section 5 the 
renormalization group equation is derived. In Section 6 we consider the example of 
$\phi^4$ and derive the above relation (\ref{result}). In Section 7 we summarize our 
findings and present a numerical estimate of the scale at which the unnatural relation 
between propagator and coupling could show up.

\section{The Dominicis-Englert Theorem}

Throughout the paper we use the functional methods described in 
the textbook \cite{Vasiliev:1998cq}, where a discussion of the Dominicis-Englert theorem can be found. 

Let the generating functional of connected renormalized Green functions $W(J)$ be related to the 
bare action of the theory $I_B$:
\begin{equation} 
\label{i_b}
e^{W(J)}=\int\mathcal{D}\phi\; e^{I_B(\phi) + J\phi}.
\end{equation}
$I_B$ diverges when ultraviolet regularization is removed, while $W(J)$ stays finite. For simplicity, 
we consider only the case of commuting fields. Generalization to the case of anticommuting fields 
is straightforward.

Then the Dominicis-Englert theorem states that
\begin{equation}
\label{det}
e^{I_B(\phi)}=\int\mathcal{D}J\; e^{W(J) - J\phi}.
\end{equation}
This is the duality: consider $W(J)$ as action, compute its connected Green functions and 
their generating functional will coincide with $I_B(-\phi)$.

It is easy to formally check this statement. Indeed, substitute into (\ref{det}) the standard 
perturbative representation of $W$, $\exp(W(J)) = \exp(V_B(\delta/\delta J))\exp(JD_BJ/2)$, 
where $D_B$ is the bare propagator,
$I_B(\phi) = -\phi D_B^{-1}\phi/2 + V_B(\phi)$:
\begin{eqnarray}
\label{detder}
e^{I_B(\phi)}&=&\int\mathcal{D}J\; e^{ - J\phi}e^{V_B(\delta/\delta J)}e^{JD_BJ/2}\nonumber\\
&=&e^{V_B(\phi)}\int\mathcal{D}J\; e^{ - J\phi}e^{JD_BJ/2},
\end{eqnarray}
where the last line was obtained via functional integration by parts (functional derivatives 
in $J$ were moved on the source term). Finally, the Gaussian integration in the last line proves 
the statement.

Let us now separate the part quadratic in the sources in the right-hand-side of (\ref{det}):
\begin{eqnarray}
\label{sep}
e^{I_B(\phi)}&=&e^{\bar{W}(-\delta/\delta \phi)}\int\mathcal{D}J\; e^{JDJ/2 - J\phi}\nonumber\\
&=&e^{\bar{W}(-\delta/\delta \phi)}e^{-\phi R\phi/2},
\end{eqnarray}
where $D$ is the renormalized propagator: $W(J)=JDJ/2+\bar{W}(J)$, $\bar{W}$ does not contain any part 
quadratic in the source, and $R\equiv D^{-1}$. Notice, that if $\bar{W}$ contains a term linear in the 
source (i.e., the field has a nontrivial vacuum expectation) it can be accounted for via shifting the 
field by the vacuum expectation. So in what follows we assume that expansion of $\bar{W}$ in powers 
of sources starts from a cubic term, and the field is counted from its vacuum value.

The right-hand-side of (\ref{sep}) contains Feynman integrals with $R$ as their propagator, and $\bar{W}$ 
as the generating functional for the vertexes. Clearly, it can be rewritten in a more familiar form 
with $D$ as the propagator. This is achieved with a sequence of standard tricks (see \cite{Vasiliev:1998cq}). 
First we exchange the places of the propagator and vertices in the above formula:
\begin{equation}
\label{trick}
e^{\bar{W}(-\delta/\delta \phi)}e^{-\phi R\phi/2} = 
\big(e^{-\frac{\delta}{\delta J}\frac{R}{2}\frac{\delta}{\delta J}}
e^{\bar{W}(J)-J\phi}\big)_{J=0}.
\end{equation}
In the right-hand-side of this equation, the variational derivatives act on a product of two 
exponentials. Thus, to compute the right-hand side, we can introduce an independent source 
in each exponential, replace each derivative with a sum of two derivatives, and, after 
acting with derivatives, set back the two sources to a common (zero) value:
\begin{widetext}
\begin{eqnarray}
\label{trick2}
\big(e^{-\frac{\delta}{\delta J}\frac{R}{2}\frac{\delta}{\delta J}}
e^{\bar{W}(J)-J\phi}\big)_{J=0}&=&\big(e^{-(\frac{\delta}{\delta J_1}+\frac{\delta}{\delta J_2})\frac{R}{2}
(\frac{\delta}{\delta J_1}+\frac{\delta}{\delta J_2})}
e^{\bar{W}(J_1)-J_2\phi}\big)_{J_1=J_2=0}\nonumber\\
&=&e^{-\phi\frac{R}{2}\phi}\big(e^{(R\phi)\frac{\delta}{\delta J_1}}
\left[e^{-\frac{\delta}{\delta J_1}\frac{R}{2}\frac{\delta}{\delta J_1}}e^{\bar{W}(J_1)}\right]\big)_{J_1=0}\nonumber\\
&=&e^{-\phi\frac{R}{2}\phi}\left[e^{-\frac{\delta}{\delta J_1}\frac{R}{2}\frac{\delta}{\delta J_1}}e^{\bar{W}(J_1)}\right]_{J_1=R\phi}.
\end{eqnarray}
\end{widetext}
Here the last line is obtained by noticing that the exponential of the variational derivatives 
acting on the square bracket shifts the argument $J_1$. The last step in this chain of tricks 
is to switch over in the square bracket from the variable $J_1$ to the variable $\phi=DJ_1$. 
Recalling that we were performing a transformation of the right-hand-side of (\ref{det}), 
we obtain the desired form of this relation:
\begin{equation}
\label{det1}
e^{I_B(\phi)} = 
e^{-\phi\frac{R}{2}\phi}e^{-\frac{\delta}{\delta \phi}\frac{D}{2}\frac{\delta}{\delta \phi}}e^{V(\phi)},
\end{equation}
where $V(\phi)\equiv \bar{W}(R\phi)$ is the generating functional of vertex functions, which are 
renormalized connected Green functions with amputated external propagators (because each source 
in $\bar{W}(J)$ is replaced with $R\phi$, where $R$ is the inverse propagator). Expanding the 
right-hand-side of this equation in powers of $D$ generates Feynman diagrams. Each $V$ 
corresponds to a vertex, each $-(\delta/\delta \phi)\frac{D}{2}(\delta/\delta \phi)$ 
corresponds to a line of a diagram. 

Let us take logarithm of both sides of (\ref{det1}):
\begin{equation}
\label{detf}
I_B(\phi)=- \phi\frac{R}{2}\phi + T_c\; e^{V(\phi)}.
\end{equation}
Here $T_c$ is the operation that differs 
$\exp\big(-(\delta/\delta \phi)\frac{D}{2}(\delta/\delta \phi)\big)$ only by suppression of 
the disconnected Feynman diagrams:
\begin{equation}
\label{tc}
T_c V^n(\phi) \equiv \left[e^{-\frac{\delta}{\delta \phi}\frac{D}{2}\frac{\delta}{\delta \phi}} V^n(\phi)\right]_c,
\end{equation}
where the subscript $c$ by the square bracket means that only connected diagrams are kept 
within brackets.

This relation is correct because it is known \cite{Vasiliev:1998cq} that taking logarithm of a 
generating functional of an $S$-matrix amounts to discarding  disconnected Feynman diagrams. 

We point out that an equation for the Wilsonian effective action resembling relation (\ref{detf}) 
has been previously introduced in \cite{Rosten:2008ts}.

Relation (\ref{detf}) is the form of the Dominicis-Englert theorem which we will use.

\section{The Inaction Equations and the Normalized Action}

For a perturbatively renormalizable Poincare invariant theory, the left-hand-side of equation (\ref{detf}) 
is a local Poincare invariant functional of fields with no couplings of negative mass dimensions. 
Such functionals form a finite dimensional linear subspace in the linear space of all functionals. 
Let $P_\mu$ be a projector onto this subspace from the space of all functionals. In particular, 
this means that for the theories with scalar, spinor and vector fields action of $P_\mu$ on any 
functional of fields yields a local polynomial in fields of not more than fourth power. The subscript $\mu$ 
denotes a generalized normalization point. Obviously, a space of projectors from an infinite-dimensional 
space to a finite-dimensional subspace is infinite-dimensional. For practical purposes, it is needed 
to fix a finite dimensional smooth surface in the space of all such projectors. Then $\mu$ denotes the 
coordinates on this surface. The arbitrariness of choice here corresponds to arbitrariness of fixing 
the renormalization scheme. Below (Section 6) we will fix the renormalization scheme in a particular way. 
For now let us consider the most general formulation. We will use the projector properties, 
$P_\mu I_B = I_B$ and $P_\mu^2 = P_\mu$. 

Act on both sides of (\ref{detf}) by $(1-P_\mu)$. The result is an equation for the renormalized 
propagator $D\equiv R^{-1}$ and for the generating functional of the renormalized connected vertecies $V$:
\begin{equation}
\label{inaction}
0 = (1-P_\mu)\left[- \phi\frac{R}{2}\phi + T_c\; e^{V(\phi)}\right].
\end{equation}

We call this \textit{inaction equation}, because it does not involve the bare action. This equation can 
be used to determine Green functions of the theory in a way avoiding ultraviolet infinities.

To set up a scheme of solving (\ref{inaction}) for $V$ and $R$, let us single out the desired $V$ in its 
right-hand-side:
\begin{widetext}
\begin{eqnarray}
\label{V}
0&=& (1-P_\mu)\left[- \phi\frac{R}{2}\phi + V(\phi) +(T-1)_c\; e^{V(\phi)}\right]\nonumber\\
&=&- \phi\frac{R}{2}\phi + V(\phi) - I_\mu(\phi) + (1-P_\mu)(T -1)_c\; e^{V(\phi)},
\end{eqnarray}
\end{widetext}
where the operation $(T-1)_c$ differs the above $T_c$ by suppressing the diagram without lines.
Here, in the first line, we noticed that the only connected diagram without lines is the vertex $V$ itself, 
and, in the second line, a new object is introduced:
\begin{equation}
\label{na}
I_\mu(\phi)\equiv P_\mu\left[- \phi\frac{R}{2}\phi + V(\phi)\right].
\end{equation}
We call it \textit{normalized action}. Finally, the equation for the renormalized connected Green 
functions which we will solve is obtained from (\ref{V}) by taking a part of its right-hand-side to the left:
\begin{equation}
\label{equation}
- \phi\frac{R}{2}\phi + V =  I_\mu(\phi) - {\mathcal T} \; e^{V(\phi)},
\end{equation}
where we introduced a new operation 
\begin{equation}
\label{tprod}
{\mathcal T}\equiv (1-P_\mu)(T -1)_c. 
\end{equation}
We will call this operation \textit{renormalized $T$-product}.

We introduced above two important objects: the normalized action and the renormalized $T$-product. Let us 
list some of their properties. As follows from the definition (\ref{na}), the normalized action is a local 
functional with no couplings of negative mass dimensions. In other words, it is similar to the bare action 
with the only crucial distinction: it is \textit{finite}. This is the case simply because it is the local part 
of the finite $\left[-\phi \frac{R}{2} \phi + V(\phi)\right]$ extracted from it via the action of $P_\mu$. 
The renormalized $T$-product is also very similar to the ordinary $T$-product. The difference is that, first, 
it skips disconnected diagrams, and, second, after forming the Feynman diagrams the renormalized $T$-product 
drops the local part of them via the action of $(1-P_\mu)$. Again, the crucial difference with respect to 
the ordinary $T$-product is that the outcome of the action of the renormalized $T$-product is finite if it 
acts on a product of vertex functionals $V$ of a renormalizable theory. 

We further define the normalized propagator $D_\mu \equiv R_\mu^{-1}$ and the normalized vertex functional 
$V_\mu(\phi)$:
\begin{eqnarray}
\label{decomp}
-\phi\frac{R_\mu}{2}\phi &\equiv& P_\mu\left[-\phi\frac{R}{2}\phi\right],\nonumber\\ 
V_\mu(\phi) &\equiv& P_\mu\left[V(\phi)\right],\nonumber\\
I_\mu(\phi) &=& -\phi\frac{R_\mu}{2}\phi + V_\mu(\phi).
\end{eqnarray}
Here the separation between the quadratic part involving $R_\mu$ and $V_\mu$ is supported by the condition 
that the expansion of $V_\mu$ in powers of the fields starts from a cubic term.

In the next Section we will work out the expansion of the propagator $D$ and the vertex functional $V$ in 
powers of the normalized vertex functional $V_\mu(\phi)$.

\section{The Finite Perturbation Theory}

Let us assume that $I_\mu(\phi)$ is given and try to solve equation (\ref{equation}) for $R$ and $V(\phi)$. 
We may do it by iterations taking as the starting point $I_\mu(\phi)$ (which means that we start with $D_\mu$ 
as $D$ and $V_\mu(\phi)$ as $V(\phi)$). Computing the right-hand-side of (\ref{equation}) yields the first 
iteration for $D$ and $V$, we plug it again to the right-hand-side to obtain the second iteration, etc. 
This procedure generates expansions of $D$ and $V$ in powers of $V_\mu$. No infinities may appear in this 
process since we are expanding finite quantities in powers of finite quantities. (Of course, this is only 
the case if we are dealing with a renormalizable theory.) It is useful to keep in mind an analogy with 
the standard perturbation theory. Within this analogy, $D_\mu$ plays the role of the free propagator, 
and $V_\mu(\phi)$ represents the collection of vertexes of the perturbation theory. 

The normalized vertex functional $V_\mu$ contains couplings of the theory renormilized within a particular renormalization scheme. They can be expanded in powers of the couplings renormalized in some other scheme 
(e.g., in the minimal subtraction scheme). Withing perturbation theory, there is a one-to-one mapping
between the couplings renormalized in different schemes \cite{Pivovarov:1988ej}. Thus, one can reproduce 
the standard perturbation theory expanding the genrating functional of the Green functions in powers of the
functional $V_\mu$. 

It is possible to give a closed expression for the $n$-th order of this expansion in terms of the preceding 
orders. To this end, we first need to single out from the renormalized $T$-product involved in (\ref{equation}) 
the zeroth order part. Recall that ${\mathcal T}$ involves the propagator $D$ (see equations (\ref{tprod}) 
and (\ref{tc})). Its inverse $R$ contains the zeroth order piece $R_\mu$ and the rest, $\Sigma$:
\begin{equation}
\label{dd}
R\equiv R_\mu + \Sigma,
\end{equation}
where $\Sigma$ (an analog of the self-energy) has an expansion in positive powers of $V_\mu$ 
(i.e., it vanishes with $V_\mu$). 
Obviously, we can replace the propagator involved in ${\mathcal T}$ with the analog of the free propagator 
$D_\mu\equiv R_\mu^{-1}$. The change in the propagator can be compensated via adding a new vertex 
$\phi\frac{\Sigma}{2}\phi$ to the vertex functional $V(\phi)$:
\begin{equation}
\label{freeprop}
{\mathcal T} e^{V(\phi)} = {\mathcal T}_\mu e^{V(\phi) + \phi\frac{\Sigma}{2}\phi},
\end{equation}
where
\begin{equation}
\label{tmu}
{\mathcal T}_\mu \equiv (1-P_\mu)\left[e^{-\frac{\delta}{\delta \phi}\frac{D_\mu}{2}\frac{\delta}{\delta \phi}} -1 \right]_c.
\end{equation}
As before, the subscript $c$ by the square bracket means that the disconnected diagrams appearing after 
the action of the variational derivatives must be discarded.

With equations (\ref{tmu}), (\ref{dd}), and (\ref{decomp}), the equation (\ref{equation}) can be rewritten as follows:
\begin{equation}
\label{transeq}
-\phi\frac{\Sigma}{2}\phi + V(\phi) = V_\mu(\phi) - {\mathcal T}_\mu e^{V(\phi) + \phi\frac{\Sigma}{2}\phi}.
\end{equation}

The next step in transforming equation (\ref{equation}) to a form suitable for constructing the perturbation theory 
consists in singling out from the exponential in the right-hand-side of (\ref{transeq}) the part linear in 
$\Sigma$ and $V$, and in moving this linear part to the left-hand-side:
\begin{widetext}
\begin{equation}
\label{nexteq}
(1+{\mathcal T}_\mu)\left[-\phi\frac{\Sigma}{2}\phi + V(\phi)\right] =  
V_\mu(\phi) - {\mathcal T}_\mu \left(e^{V(\phi) + \phi\frac{\Sigma}{2}\phi} - 1 - V(\phi) -\phi\frac{\Sigma}{2}\phi\right).
\end{equation}
\end{widetext}
Here we have used the observation that ${\mathcal T}_\mu$ nullifies any constant and any functional quadratic in 
the fields. The constant is nullified because ${\mathcal T}_\mu$ contains at least two variational derivatives 
(i.e., it generates at list one line of a Feynman diagram), and any quadratic functional is nullified because 
the action of the variational derivatives involved in ${\mathcal T}_\mu$ makes it a constant independent of 
fields, which is a local functional, and the definition of ${\mathcal T}_\mu$ contains as the leftest factor 
the operation $(1-P_\mu)$ projecting out the local functionals (see definition of ${\mathcal T}_\mu$ in (\ref{tmu})).

The last step in this chain of transformations of (\ref{equation}) is to act on both sides of (\ref{nexteq}) 
with the operation $(1+{\mathcal T}_\mu)^{-1}$:
\begin{widetext}
\begin{equation}
\label{finaleq}
-\phi\frac{\Sigma}{2}\phi + V(\phi) =  
V_\mu(\phi) - (1+{\mathcal T}_\mu)^{-1}{\mathcal T}_\mu \left(e^{V(\phi) + 
\phi\frac{\Sigma}{2}\phi} - 1 - V(\phi) -\phi\frac{\Sigma}{2}\phi\right).
\end{equation}
\end{widetext}
Here we used the fact that ${\mathcal T}_\mu$ nullifies $V_\mu$, because the latter is a local functional. 
The inverse of $(1 + {\mathcal T}_\mu)$ is understood as the geometric series in powers of ${\mathcal T}_\mu$. 
This equation is convenient because the expansion of its right-hand-side in powers of $\Sigma$ and $V_\mu$ 
starts from the second power.

We will solve (\ref{finaleq}) for $\Sigma$ and $V$ looking for solutions as power series in 
$V_\mu$ that start from a term linear in $V_\mu$: 
\begin{eqnarray}
\label{pseries}
\Sigma &=& \sum_{k=1}^\infty \Sigma_k,\nonumber\\
V(\phi) &=& \sum_{k=1}^\infty V_k(\phi),
\end{eqnarray}
where the terms with the subscript $k$ are of order $(V_\mu)^k(\phi)$.

Substituting these expansions in (\ref{finaleq}) and equaling terms of the same order in both sides we 
obtain a collection of equations defining $\Sigma_k$ and $V_k(\phi)$. For example, in the leading order we have
\begin{equation}
\label{leading}
-\phi\frac{\Sigma_1}{2}\phi + V_1(\phi) =  V_\mu(\phi).
\end{equation}
This is the case because the exponential in the right-hand-side of (\ref{finaleq}) does not contribute 
to the leading order. Since the expansion of $V_\mu$ in powers of fields starts from a cubic term, 
this relation implies that $\Sigma_1=0$, and $V_1(\phi) = V_\mu(\phi)$.

As soon as we defined the leading terms in the expansions (\ref{pseries}), equation (\ref{finaleq}) 
defines unambiguously the higher orders of the expansions. Indeed, computation of the exponential in 
the right-hand-side of (\ref{finaleq}) at the order $(V_\mu)^k$ requires the knowledge of 
$\Sigma_l$ and $V_l$ at $l<k$. In turn, computing the exponential at the order $k$ provides us 
with the knowledge of $\Sigma_k$ and $V_k$.

For example,
\begin{equation}
\label{second}
-\phi\frac{\Sigma_2}{2}\phi + V_2(\phi) = - (1+{\mathcal T}_\mu)^{-1}{\mathcal T}_\mu 
\left(\frac{V_\mu(\phi)^2}{2}\right).
\end{equation}
We will not need the terms higher than this second order in this paper. But let us take a closer 
look at this second order approximation.  

First of all, the diagrams generated by the action of powers of ${\mathcal T}_\mu$ on the 
square of $V_\mu$ in the right-hand-side of (\ref{second}) are connected diagrams of two vertexes, 
each vertex has not more than four adjacent lines (recall that $V_\mu = P_\mu V$) . Therefore, 
not more than three lines can appear. (A diagram with four lines is a constant and is nullified by 
$(1-P_\mu)$ of the leftest ${\mathcal T}_\mu$.) As each ${\mathcal T}_\mu$ generates not less than 
one line, any power of 
${\mathcal T}_\mu$ larger than 3 gives a zero contribution. Thus,
\begin{equation}
\label{power}
-\phi\frac{\Sigma_2}{2}\phi + V_2(\phi) = 
- \left({\mathcal T}_\mu - {\mathcal T}_\mu^{2}+{\mathcal T}_\mu^3\right) 
\left(\frac{V_\mu(\phi)^2}{2}\right).
\end{equation}

On the next step we need to resolve separate lines inside ${\mathcal T}_\mu$:
\begin{eqnarray}
\label{lines}
{\mathcal T}_\mu &=& (1-P_\mu)\left[e^l-1\right]_c,\nonumber\\
l&\equiv& -\frac{\delta}{\delta\phi}\frac{D_\mu}{2}\frac{\delta}{\delta\phi},
\end{eqnarray}
where in the second line we defined an operation $l$ which action on a product 
of vertex functionals joins vertexes with a line in every possible way.

Let us demonstrate the specification we want to make in the right-hand side of (\ref{power}) 
on the example of the second term involving the second power of ${\mathcal T}_\mu$:
\begin{equation}
\label{spec1}
{\mathcal T}_\mu^2 V_\mu^2 = (1-P_\mu)\left[e^l-1\right]_c(1-P_\mu)\left[e^l-1\right]_c V_\mu^2.
\end{equation}

We immediately notice that the subscript $c$ on the left pair of square brackets can be omitted, 
because the right pair of square brackets with the subscript $c$ yields a set of connected diagrams, 
and adding more lines cannot make any of these diagrams disconnected. Less obvious is that we can 
drop $(1-P_\mu)$ sandwiched between the left and the right pairs of the square brackets. This is 
the case because the action of $P_\mu$ belonging to this bracket yields a local functional. This functional 
stays local after the left square bracket acts on it (this is the case because joining two legs attached two a local vertex with a propagator results in a local vertex containing as a numerical factor the propagator taken at coinciding coordinates). Finally, this local functional 
is nullified by the left $(1-P_\mu)$. Also, we can expand both exponentials in the right-hand-side 
of (\ref{spec1}) keeping only the terms with not more than three lines:
\begin{equation}
\label{spec2}
{\mathcal T}^2_\mu V_\mu^2 = (1-P_\mu)\left(l\left[l +\frac{l^2}{2}\right]_c + \frac{l^2}{2}
\left[l\right]_c\right) V_\mu^2.
\end{equation}

Next step in this chain of specifications of the formula (\ref{second}) is to label differently 
the field arguments in the two vertexes, and replace each variational derivative within each 
line generating operation $l$ with a sum of two derivatives. This resolves each line into a 
sum of three terms: one term joins two different vertexes, another joins vertex number one 
with itself, and the third term joins vertex number two with itself. For example,
\begin{widetext}
\begin{eqnarray}
\label{connective}
\left[l^2\right]_c V_\mu^2(\phi) 
&=& \left[\left(l_{12}^2 + 2 l_{12}(l_1+l_2)\right) 
V_\mu(\phi_1)V_\mu(\phi_2)\right]_{\phi_1=\phi_2=\phi},\nonumber\\
l_i &\equiv& -\frac{\delta}{\delta\phi_i}\frac{D_\mu}{2}\frac{\delta}{\delta\phi_i},\nonumber\\
l_{12}&\equiv& -2\frac{\delta}{\delta\phi_1}\frac{D_\mu}{2}\frac{\delta}{\delta\phi_2}.
\end{eqnarray}
\end{widetext}
Notice that we dropped the term without $l_{12}$ in the right-hand-side of the first line, 
because it would not contain a line connecting the two vertexes, and, therefore, would 
correspond to disconnected diagrams.

If we do this kind of specification with all three terms in the right-hand-side of (\ref{power}) , 
we notice a nice feature: all the terms involving powers of $l_1$ and $l_2$ cancel against 
each other, and we finally obtain
\begin{widetext}
\begin{equation}
\label{secondspecfd}
-\phi\frac{\Sigma_2}{2}\phi + V_2(\phi) = 
-\left(1-P_\mu\right)\left[\left(l_{12} - \frac{1}{2}l_{12}^2 + 
\frac{1}{3!}l_{12}^3\right)\frac{V_\mu(\phi_1)V_\mu(\phi_2)}{2}\right]_{\phi_1=\phi_2=\phi}
\end{equation}
\end{widetext} 
We remind that $V_2$ above can be separated in the right-hand-side by considering only the
contributions with more than two external legs (and, therefore, the term in the right-hand-side
involving $l_{12}^3$ does not contribute to $V_2$).

This is a nice result: within the new finite perturbation theory, the second order 
approximation contains only three diagrams, with one, two, and three lines. We 
demonstrated it for any renormalizable theory. All three diagrams are of the 
sunset topology. Diagrams of this topology are well studied \cite{Groote:2005ay}. 
In particular, it is known that there is no need to subtract subdivergencies to make 
such diagrams finite, which is in accord with the fact that only a single operation 
$(1-P_\mu)$ is present in the right-hand-side of (\ref{secondspecfd}). We point out 
that the new finite perturbation theory contains less diagrams than the conventional 
perturbation theory. We also point out the appearance of certain rational coefficients 
by the specific diagrams. For example, the diagram with three lines comes with the 
weight $(1/3 !)$. It is an interesting and important problem to extend these observations 
to higher orders of the new finite perturbation theory. 

For the present paper, it will suffice to use (\ref{secondspecfd}) for the second 
order of the finite perturbation theory.

\section{The Renormalization Group Equations}

As we have seen in the previous section, the normalized action equation (\ref{equation}) 
suffices to determine the propagator and the vertex functional. Still, the above scheme 
is not quite independent from the original formulation involving the bare action. In 
this section we augment the finite perturbation theory with a renormalization group 
equation for the normalized action. Any normalized action satisfying this equation yields  
(via the finite perturbation theory of the previous Section) propagator and vertex 
functional independent of the normalization point. So, the combination of the normalized 
action (or, equivalently, inaction) equation, finite perturbation theory, and
the renormalization group equation constitute an independent scheme of constructing 
(perturbative) quantum field theory.

To derive the renormalization group equation, let us take a derivative in $\mu$ of (\ref{equation}), 
and take into account that inverse propagator $R$ and vertex functional $V$ do not depend on $\mu$. 
This is the case because they can be constructed with the bare action $I_B$ that does not know 
anything about $\mu$. Therefore, the only $\mu$-dependent objects in (\ref{equation}) are the 
normalized action $I_\mu$ and the operation ${\mathcal T}$. The definition of the letter 
(see (\ref{tprod})) involves projector $P_\mu$ depending on $\mu$. 

Taking all this into account we derive from (\ref{equation}) the following evolution equation 
for the normalized action:
\begin{equation}
\label{rge}
\frac{\partial I_\mu(\phi)}{\partial \mu} = 
- \left[\frac{\partial P_\mu}{\partial \mu}\right]\; (T-1)_c\;e^{V(\phi)}.
\end{equation}
The operation $(T-1)_c$ is introduced in (\ref{V}).

This is indeed an evolution equation, because, as we have seen in the previous Section, 
its right-hand-side can be expressed in terms of $I_\mu$. So, at a given $I_\mu$, we can 
compute its derivative in $\mu$. There is a subtlety here: in the previous Section, we 
constructed a perturbation theory for the objects in the right-hand-side of the above 
renormalization group equation. It is quite possible, and, indeed, the case, that a sum 
of several terms of the perturbation theory for the objects in the right-hand-side, 
viz. $T_c$ and and $V$, will depend on $\mu$. This does not mean that we should include 
in the right-hand-side the derivatives of these objects in $\mu$, because their dependence 
on $\mu$ is only a drawback of the approximation.

By its form the renormalization group equation (\ref{rge}) resembles the so-called exact
renormalization group equation of \cite{Polchinski:1983gv}. In this equation, the derivative in
the cutoff of an effective action is expressed in terms of the effective action. An important 
distinction of our approach is that it does not evoke any cutoff (and any regularization). 
Instead, the derivative in our 
equation is taken in the normalization point. This may be important for applications of our approach 
to models with gauge symmetries. As known, gauge symmetries do not go well with cutoff. Because of
this, application of the exact renormalization group equation to gauge theories is quite involved
\cite{Arnone:2005fb}. Application of our approach to gauge theories is beyond the scope of this paper, 
but we expect that our approach will be almost directly applicable to gauge theories. At the same time, the exact renormalization group equation allows for nonperturbative considerations. A number of important 
nonperturbative notions has been introduced within the approach of exact renormalization group (for a recent
review see \cite{Rosten:2010vm}). We speculate that incorporating some of these nonperturbative notions into the
approach of this paper may require considering projectors $P_\mu$ qualitatevely different from the ones considered 
in this paper.

The single equation (\ref{rge}) is in fact a collection of equations for the couplings 
involved in $I_\mu$. (And we also can consider cases where $\mu$ is not one-dimensional; 
in such cases the derivative over $\mu$ should be replaced with the derivative over 
the coordinates of the point $\mu$.) As we will see in the next section, these couplings 
are in one-to-one correspondence with the Green functions of not more than four fields. 
A set of equations can be derived from (\ref{rge}) that describes the dependence of 
these Green functions on the momenta. (The derivation requires a specification of 
the projector $P_\mu$ and is relegated to the next Section.) In the following, 
the terms ``renormalization group equation'' and ``renormalization group equations'' 
can refer either to equation (\ref{rge}), or to the equations for the Green functions 
derived from (\ref{rge}). Which meaning is used is always clear from the context.  

Overall, we suggest the following scheme of computations in quantum field theory. The 
propagators and the amplitudes involving not more than four particles are determined 
via the renormalization group equations, and the rest of the amplitudes are determined 
via the finite perturbation theory of the previous Section. 

The purpose of the next Section is to start this program for the case of $\phi^4$.

\section{The Case of $\phi^4$}

To start, we specify a projector $P_\mu$. Let the projector nullify any product of 
more than four fields. By definition, functionals we deal with within $\phi^4$ are 
sums of even powers of fields. Therefore, to determine the projector completely 
we need to specify its action on the functionals of the following form:
\begin{widetext}
\begin{eqnarray}
\label{funct}
F(\phi) &=& \frac{1}{2}\int \; dp \;\phi(p)\;\phi(-p)(2\pi)^4\left[-R(p^2)\right]\nonumber\\
        &+& \frac{1}{24}\int \; \prod_{i=1}^4 \left(dp_i\; \phi(p_i)\right) 
\delta\left(\sum_{i=1}^4 p_i\right)\;(2\pi)^4 \left[g(\{p_ip_j\})\right].
\end{eqnarray}
\end{widetext}         
The fields here are in the momentum representation. The argument of $g$ is the list of six 
scalar products of the four momenta involved, $p_ip_j$, $i<j$, $i$ and $j$ range from 1 to 4. 
Here $R$ and $g$ are arbitrary functions of, respectively, one and 
six arguments. Later on $R$ will denote the inverse propagator and $g$ the Fourier transformed 
four-point connected Green function with amputated external propagators (and with the delta 
function expressing the conservation of momentum removed).

Overall, $R$ and $g$ in (\ref{funct}) depend on seven arguments. According to this, let us 
take that $\mu$ is a
point in a seven-dimensional space, 
\begin{equation}
\label{mu}
\mu = \{Q^2,\{Q_iQ_j,1\leq i < j\leq 4\}\}. 
\end{equation}
And let the action of $P_\mu$ on $F$ from (\ref{funct}) be defined as 
follows:
\begin{widetext}
\begin{eqnarray}
\label{pmu}
P_\mu F(\phi)   &=& \frac{1}{2}\int \; dp \;\phi(p)\;\phi(-p)
(2\pi)^4\left[-(p^2-Q^2)R'(Q^2) - R(Q^2) \right]\nonumber\\
        &+& \frac{1}{24}\int \; \prod_{i=1}^4 \left(dp_i\; 
\phi(p_i)\right) \delta\left(\sum_{i=1}^4 p_i\right)\;(2\pi)^4 \left[g(\{Q_iQ_j\})\right].
\end{eqnarray}
\end{widetext}
We see that the action of $P_\mu$ had replaced the square brackets in (\ref{funct}) with 
the pieces of their Taylor expansions around the points specified by $\mu$: $R$ is expanded 
around $Q^2$ up to the first order term (here and below primes on $R$ mean derivatives in $Q^2$), 
and $g$ is expanded around $\{Q_iQ_j\}$ up to the zeroth order term.

Evolution of the normalized action in the variables $\{Q_iQ_j\}$ has been briefly considered in \cite{Pivovarov}. 
In particular, this consideration reproduces the one-loop $\beta$-function of $\phi^4$.

In this paper we will study only the evolution of the normalized action in $Q^2$. We will need 
the derivative of $P_\mu$ in $Q^2$:
\begin{widetext}
\begin{equation}
\label{pder}
\frac{\partial P_\mu}{\partial Q^2} \; F(\phi) = 
\frac{1}{2}\int \; dp \;\phi(p)\;\phi(-p)(2\pi)^4\left[-(p^2-Q^2)R''(Q^2) \right].
\end{equation}
\end{widetext}        
 
After these preparations, let us write down one of the equations (\ref{rge}) describing the evolution 
of the normalized action in the first coordinate of $\mu$, which is $Q^2$ in the notations of (\ref{mu}). 
Let us first relate the left-hand-side of (\ref{rge}) to derivatives of Green functions in momenta. 
If we recall the definition (\ref{na}), and the above formula for the action of the derivative of 
$P_\mu$ on the arbitrary functional $F(\phi)$, we obtain that
\begin{equation}
\label{lhs}
\frac{\partial I_\mu}{\partial Q^2} = \frac{1}{2}\int \; dp \;\phi(p)\;\phi(-p)(2\pi)^4\left[-(p^2-Q^2)R''(Q^2) \right],
\end{equation}        
where $R$  is now the inverse propagator of the $\phi^4$ theory.

Let us now calculate the leading order of the right-hand-side of (\ref{rge}) for the case at hand. 
First of all, $V_\mu$ coincides with the second line of (\ref{pmu}), where it is understood that 
$g(\{Q_iQ_j\})$ is the Fourier transformed connected four-point Green function with delta-function 
expressing the momentum conservation omitted and external propagators amputated. The dependence 
of $g$ on the momenta is a higher order effect and we will treat it in what follows as a (small) 
constant. 

In principle, there could be a term linear in $V_\mu$ in the right-hand-side of (\ref{rge}). 
But we notice that the action of the derivative of $P_\mu$ nullifies it since this linear 
term does not depend on $Q^2$. Thus, the expansion of the right-hand-side of (\ref{rge}) 
starts from the term quadratic in $V_\mu$ (or, equivalently, quadratic in $g$).

To compute this term, we need the second order approximation $V_2$ determined in (\ref{secondspecfd}), because 
we can make in the second order approximation the following replacement in the right-hand-side of (\ref{rge}):
\begin{equation}
\label{explanation}
(T -1)_c \; e^V \rightarrow (T-1)_c\frac{V_\mu^2}{2} + (T - 1)\; V_2,
\end{equation}
where we omitted 
subscript $c$ in the second term, because $V_2$ is already connected. 

The $T$-product in (\ref{explanation}) involves  the propagator $D$. But within the leading 
approximation it can be replaced with $D_\mu$,
\begin{eqnarray}
\label{propagator}
D_\mu (p) &=& \frac{1}{(p^2-Q^2)R'(Q^2) + R(Q^2)}\nonumber\\
&=& \frac{1}{R'(Q^2)}\times\frac{1}{p^2+M^2(Q^2)},
\end{eqnarray}
where we have introduced the notation $M^2(Q^2)\equiv R(Q^2)/R'(Q^2) - Q^2$.

Now we substitute $V_2$ from (\ref{secondspecfd}), and, after some hassle (in particular, resolving 
different types of lines, like we did in the derivation of (\ref{secondspecfd})) we obtain that 
the right-hand-side of (\ref{rge}) is
\begin{equation}
\label{rhs}
-\frac{\partial P_\mu}{\partial Q^2} \frac{1}{6}
\left[l_{12}^3 \frac{V_\mu(\phi_1)V_\mu(\phi_2)}{2}\right]_{\phi_1=\phi_2=\phi},
\end{equation}
where $l_{12}$ was defined in (\ref{connective}).

As before, all the diagrams containing lines starting and ending at the same vertex had canceled 
against each other, and we see that the right-hand-side of (\ref{rge}) in the leading order 
contains (under the action of the derivative of the projector $P_\mu$) only a single diagram 
consisting of two vertexes with four adjacent lines and three lines joining them. 
This is the so-called sunset diagram.

Both sides of (\ref{rge}) contain the derivative of $P_\mu$ in $Q^2$. Checking the 
formula (\ref{pder}) we see that equaling such derivatives means equaling the second 
derivatives in momentum squared of two functions. The function involved in the 
left-hand-side is the inverse propagator, and  it is the sunset diagram in the right-hand-side. 

In this way, we obtain a renormalization group equation for the scalar propagator in the following form:
\begin{widetext}
\begin{equation}
\label{in_terms_of_pi}
R'' = - \frac{g^2}{6(R')^3}\left[\left(\frac{\partial}{\partial Q^2}\right)^2\tilde{\Pi}(Q,M)\right]
_{M^2 = R(Q^2)/R'(Q^2) - Q^2},
\end{equation} 
where
\begin{equation}
\label{pi}
\tilde{\Pi}(Q,M) = 2 \pi^2 \int_0^\infty\left(\frac{Qx}{2}\right)^{-1}J_1(Qx)
\left[\frac{MxK_1(Mx)}{(2\pi x)^2}\right]^3 x^3 dx
\end{equation}
\end{widetext}
is the Fourier transform of a product of three scalar propagators  with mass $M$ (represented by the 
cube of the fraction with the modified Bessel $K_1$). (The last formula is copied with minor 
modifications from \cite{Groote:2005ay}.) We clarify that $R'^3$ in the right-hand-side of (\ref{in_terms_of_pi}) 
comes from rising to the third power the $R'$ involved in the right-hand-side of (\ref{propagator}). 

The last step in the derivation of the renormalization group equation is to take the 
second derivative in $Q^2$ under the integral defining $\tilde\Pi(Q,M)$. Using the 
properties of the Bessel function (see, e.g., the Appendix in
\cite{Groote:2005ay}) the derivative can be expressed in terms of the Bessel 
function of the third order. Notice that the integral in (\ref{pi}) is 
divergent at zero $x$, but after taking the second derivative in $Q^2$ we obtain
a finite integral. All this finally results in the equation for the inverse 
propagator (\ref{prop}) advertised in the Introduction. The rest of this Section 
is devoted to a study of equation (\ref{prop}).

First of all, let us rewrite the second order differential equation (\ref{prop}) as a 
pair of coupled first order equations. As discussed in the Introduction, convenient 
variables are $m^2(Q^2)\equiv \left(R/(Q^2 R')-1\right)$ and $n(Q^2)\equiv (R')^4$. 
Also it is convenient to use as evolution parameter a $\log$ of $Q^2$, $t = \log (Q^2/M^2)$. 
It is easy to check that (\ref{prop}) is equivalent to the following pair of the 
first order equations:
\begin{eqnarray}\label{mnvar}
\frac{d}{dt}\;m^2 &=&-m^2 + \frac{\gamma_\phi}{n}\;(1 + m^2)\;\Phi(m),\\
\label{mnvar2}
\frac{d}{dt}\;n &=&-4 \gamma_\phi \;\Phi(m),	
\end{eqnarray}
where 
\begin{equation}\label{Phi}
\Phi(m) \equiv 8 \int_0^\infty J_3(x)\left[mK_1(mx)\right]^3 x\; dx
\end{equation}

This evolution system should be given the initial conditions. We take that $m^2(0)= 1$ and $n(0)=1$, 
which implies that $M$ in the definition of the evolution parameter is the physical (pole) mass.

The next step is to study the dependence of function $\Phi$ on $m$. Asymptotics of this function 
can be computed with Wolfram's \textit{Mathematica} \cite{Wolfram:2008}. At zero $m$ it has unit value, 
its asymptotic 
at large $m$ is $0.3609/(6\; m^2)$ ($0.3609$ is approximately the value of the Meijr $G$-function 
appearing in the asymptotic). For positive $m$,
\begin{equation}\label{approx}
\Phi(m) \approx \frac{0.3609}{6\; m^2 + 0.3609}.
\end{equation} 
This approximation improves with growing $m$. Its accuracy deteriorates at small $m$ because 
$\Phi$ has negative derivative in $m$ at zero $m$. Still, the error is less than 20\%, and 
the approximation again improves at very small $m$ because at zero $m$ the error vanishes. 
For our purposes, we take that $\Phi$ is unit for $m^2\leq\gamma_\phi$, and is given 
by (\ref{approx}) otherwise.

Now we are ready to study the dynamical system (\ref{mnvar}), (\ref{mnvar2}). 

First, we notice that nothing interesting happens at momenta smaller than $M$. The terms 
in the right-hand-sides of (\ref{mnvar}), (\ref{mnvar2}) involving $\gamma_\phi$ can be safely 
neglected. This is the case because $m^2$ is growing at decreasing negative $t$ and the 
derivative of $n$ is decreasing in absolute value, while $n$ itself is slightly growing 
with momentum decreasing. Numerical experiments confirm this observation. At low momenta 
inverse propagator $R(Q^2)$ is very close to the free inverse propagator $Q^2+M^2$.

Much more interesting is the situation at momentum growing larger than $M$. The key point 
is that there are two competing terms in the right-hand-side of (\ref{mnvar}). The first 
term is negative and decreasing, the second term is positive and increasing. They will 
inevitably cancel each other at a certain value of $Q^2$. At this $Q^2$ the derivative 
of mass squared in $t$ vanishes and mass stays approximately constant. This happens 
around the point 
$M^2/Q^2 = \gamma_\phi$. Here the minimal value of the running mass is reached, 
$m^2_{min} \approx \gamma_\phi$.
At larger momentum the derivative of mass becomes positive. The running mass in units 
of normalization point starts to grow in contradiction with the dimensional analysis!
The normalization factor $n$ is slowly decreasing around this point: 
$n= 1- 4\gamma_\phi \log (Q^2/M^2)$. This decrease causes extra growth in $m^2$. 
At this stage, $m^2 \approx \gamma_\phi/n$. We conclude that at this stage the 
running mass can be expressed in terms of the anomalous dimension of the field:
\begin{equation}\label{improved}
M^2(Q^2) \approx \frac{\gamma_\phi\; Q^2}{1 - 4\gamma_\phi\log (Q^2/M^2)},
\end{equation}
which is an improved version of (\ref{result}). The range of validity of this equation is 
$M^2/\gamma_\phi<Q^2\ll M^2\exp\left(1/(4\gamma_\phi)\right)$.

Next starts a transit stage when $m^2$ had grown back to values of unit order. It is not 
possible anymore to use the approximation $\Phi \approx 1$, because mass is not small enough. 
This is a rather lengthy stage because the increase in mass slows down the evolution 
($\Phi$ decreases with mass growing). The best we can do for this stage is to study it numerically.

The next stage of the evolution starts when $Q^2$ becomes larger than $M^2 \exp(1/(4\gamma_\phi) - 1/4)$. 
Here $m^2$ becomes larger than unit.  As was discussed in the Introduction, at this stage the 
perturbation theory becomes unreliable because of the smallness of $n$, and we include it for 
the sake of completeness. It is more convenient to go back to the equation (\ref{prop}) to 
describe this stage of the evolution. Indeed, if we replace the integral of the Bessel 
functions in the right-hand-side of (\ref{prop}) with its large mass asymptotic, we obtain
\begin{equation}\label{lmas}
R''(Q^2) = - \frac{\gamma_\phi}{Q^2(R')^3}\times \frac{0.3609}{6\; (R/(Q^2R')-1)}.
\end{equation}
Now, if we neglect the unit in the denominator of the second factor (this is justified because  
$m^2\gg 1$ on this stage), we obtain
\begin{equation}\label{simpl}
R'' = -\frac{\gamma_\phi\times 0.3609}{6\; (R')^3} \left(\log\frac{R}{2M^2}\right)',
\end{equation}      
where the scale in the $\log$ is chosen to make the $\log$ vanish at $Q^2=M^2$. This is equivalent to
\begin{equation}\label{integr}
R' = \left[1 - 0.3609\;\frac{2\gamma_\phi}{3} \log\frac{R}{2M^2}\right]^{\frac{1}{4}}.
\end{equation}
We see that the evolution stops at the asymptotic value 
$R_a = 2 M^2 \exp(\frac{3}{2\gamma_\phi}\times\frac{1}{0.3609})$. The asymptotic value is reached at 
a critical value of the momentum squared $Q^2_c\propto R_a/\gamma_\phi$. We computed the numeric 
coefficient by this estimate for the critical value of the momentum squared, which we skip for brevity.

To summarize, there are four stages in the evolution of the inverse propagator. 
The first is the stage of the low momentum, $Q^2\leq M^2/\gamma_\phi$. At this 
stage the exact propagator is very close to the free propagator. The second stage 
is the stage of moderate momenta, $M^2/\gamma_\phi\leq Q^2\ll M^2 \exp\left(1/(4\gamma_\phi)\right)$. 
At this stage the deviation of the propagator from the free propagator becomes quite 
sizable, and there is a qualitative difference with respect to the free theory: 
the running mass in the units of normalization point starts to grow in contradiction 
with the dimensional analysis of the free theory. At the last stage of the evolution 
the momentum is really large, $Q^2\geq M^2 \exp\left(1/(4\gamma_\phi - 1/4)\right)$. Here 
perturbation theory is unreliable, but its leading order gives a meaningful answer: 
the solution to the leading order renormalization group equation for the propagator 
becomes local in the coordinate space. In other words, the soluiton 
becomes independent of the momentum beyond a critical value of the momentum. 
There is also an intermediate stage interpolating between the second and the 
fourth stage, of which we have no much to say.

\section{Conclusions} 

We constructed a finite perturbation theory for Green functions of renormalizable 
quantum field theories. Its major feature is that it avoids infinite quantities. 
In particular, it does not employ the notion of the bare action. An integral part 
of this new formulation of the perturbative quantum field theory is the 
renormalization group equations for Green functions with not more than four 
external legs. The new renormalization group equations describe the dependence 
of these Green functions on the momenta.

Applying the new formalism to $\phi^4$ theory we discovered a new phenomenon: 
For scalar field the running mass in units of the normalization point is not 
a monotonous function of the normalization point. It has a minimum at a 
certain value of the normalization point. 

May this new phenomenon be of importance for the LHC era? Within $\phi^4$ 
the discussed phenomenon is a two-loop effect. The momentum needed to 
reach the minimum of the running mass is $\left(1/\sqrt{\gamma_\phi}\right)$ 
times larger than the physical mass of the scalar particle. If we take that 
the self-coupling of the scalar field is close to unit, we obtain that we 
need to reach the momentum which is $16\pi^2\sqrt{12}\approx 174$ times 
larger than the physical mass, which is not very promising. Within the 
standard model, there is a contribution to the anomalous dimension of the 
scalar field already in the one-loop approximation. Optimistically, 
the momentum needed to observe the unnaturalness of the scalar field 
may be only $4\pi\approx 12.6$ times larger than the scalar mass. 
We conclude that an application of the considered formalism to the 
standard model is a necessary objective.

\begin{acknowledgments}
The author thanks V.~T.~Kim, L.~N.~Lipatov, V.~A.~Matveev, and  V.~A.~Rubakov for 
helpful discussions. This work was 
supported in part by the grants RFBR 08-01-00686 and President of RF NS-1616.2008.2. 
It has been written during a visit of the author to 
CERN Theoretical Physics Unit, which the author thanks for the hospitality.
\end{acknowledgments}

\nocite{*}

\bibliography{apsrgn_3}

\end{document}